\documentclass[11pt,a4paper]{article}
\usepackage{amssymb} \usepackage{amsmath} \usepackage{graphicx}
\usepackage{epsfig,latexsym}
\begin{document}

\def\bef{\begin{figure}}
\def\eef{\end{figure}}
\newcommand{\ans}{ansatz }
\newcommand{\be}[1]{\begin{equation}\label{#1}}
\newcommand{\beq}{\begin{equation}}
\newcommand{\ee}{\end{equation}}
\newcommand{\beqn}[1]{\begin{eqnarray}\label{#1}}
\newcommand{\eeqn}{\end{eqnarray}}
\newcommand{\bd}{\begin{displaymath}}
\newcommand{\ed}{\end{displaymath}}
\newcommand{\mat}[4]{\left(\begin{array}{cc}{#1}&{#2}\\{#3}&{#4}
\end{array}\right)}
\newcommand{\matr}[9]{\left(\begin{array}{ccc}{#1}&{#2}&{#3}\\
{#4}&{#5}&{#6}\\{#7}&{#8}&{#9}\end{array}\right)}
\newcommand{\matrr}[6]{\left(\begin{array}{cc}{#1}&{#2}\\
{#3}&{#4}\\{#5}&{#6}\end{array}\right)}
\newcommand{\cvb}[3]{#1^{#2}_{#3}}
\def\lsim{\raise0.3ex\hbox{$\;<$\kern-0.75em\raise-1.1ex
e\hbox{$\sim\;$}}}
\def\gsim{\raise0.3ex\hbox{$\;>$\kern-0.75em\raise-1.1ex
\hbox{$\sim\;$}}}
\def\abs#1{\left| #1\right|}
\def\simlt{\mathrel{\lower2.5pt\vbox{\lineskip=0pt\baselineskip=0pt
           \hbox{$<$}\hbox{$\sim$}}}}
\def\simgt{\mathrel{\lower2.5pt\vbox{\lineskip=0pt\baselineskip=0pt
           \hbox{$>$}\hbox{$\sim$}}}}
\def\unity{{\hbox{1\kern-.8mm l}}}
\newcommand{\eps}{\varepsilon}
\def\ep{\epsilon}
\def\ga{\gamma}
\def\Ga{\Gamma}
\def\om{\omega}
\def\omp{{\omega^\prime}}
\def\Om{\Omega}
\def\la{\lambda}
\def\La{\Lambda}
\def\al{\alpha}
\newcommand{\ov}{\overline}
\renewcommand{\to}{\rightarrow}
\renewcommand{\vec}[1]{\mathbf{#1}}
\newcommand{\vect}[1]{\mbox{\boldmath$#1$}}
\def\tm{{\widetilde{m}}}
\def\mcirc{{\stackrel{o}{m}}}
\newcommand{\Dm}{\Delta m}
\newcommand{\dm}{\varepsilon}
\newcommand{\tanb}{\tan\beta}
\newcommand{\nbar}{\tilde{n}}
\newcommand\PM[1]{\begin{pmatrix}#1\end{pmatrix}}
\newcommand{\up}{\uparrow}
\newcommand{\down}{\downarrow}
\def\omE{\omega_{\rm Ter}}
%

\newcommand{\Dsusy}{{susy \hspace{-9.4pt} \slash}\;}
\newcommand{\DCP}{{CP \hspace{-7.4pt} \slash}\;}
\newcommand{\mc}{\mathcal}
\newcommand{\gr}{\mathbf}
\renewcommand{\to}{\rightarrow}
\newcommand{\gtc}{\mathfrak}
\newcommand{\wh}{\widehat}
\newcommand{\br}{\langle}
\newcommand{\kt}{\rangle}


\def\lsim{\mathrel{\mathop  {\hbox{\lower0.5ex\hbox{$\sim$}
\kern-0.8em\lower-0.7ex\hbox{$<$}}}}}
\def\gsim{\mathrel{\mathop  {\hbox{\lower0.5ex\hbox{$\sim$}
\kern-0.8em\lower-0.7ex\hbox{$>$}}}}}

\def\nn{\\  \nonumber}
\def\de{\partial}
\def\brf{{\mathbf f}}
\def\bbf{\bar{\bf f}}
\def\bF{{\bf F}}
\def\bbF{\bar{\bf F}}
\def\bA{{\mathbf A}}
\def\bB{{\mathbf B}}
\def\bG{{\mathbf G}}
\def\bI{{\mathbf I}}
\def\bM{{\mathbf M}}
\def\bY{{\mathbf Y}}
\def\bX{{\mathbf X}}
\def\bS{{\mathbf S}}
\def\bb{{\mathbf b}}
\def\bh{{\mathbf h}}
\def\bg{{\mathbf g}}
\def\bla{{\mathbf \la}}
\def\bmu{\mathbf m }
\def\by{{\mathbf y}}
\def\bmu{\mbox{\boldmath $\mu$} }
\def\bsig{\mbox{\boldmath $\sigma$} }
\def\bunity{{\mathbf 1}}
\def\cA{{\cal A}}
\def\cB{{\cal B}}
\def\cC{{\cal C}}
\def\cD{{\cal D}}
\def\cF{{\cal F}}
\def\cG{{\cal G}}
\def\cH{{\cal H}}
\def\cI{{\cal I}}
\def\cL{{\cal L}}
\def\cN{{\cal N}}
\def\cM{{\cal M}}
\def\cO{{\cal O}}
\def\cR{{\cal R}}
\def\cS{{\cal S}}
\def\cT{{\cal T}}
\def\eV{{\rm eV}}
%




\large
 \begin{center}
 {\Large \bf Quantum chaos inside space-temporal Sinai billiards}
 \end{center}

 \vspace{0.1cm}

 \vspace{0.1cm}
 \begin{center}
{\large Andrea Addazi}\footnote{E-mail: \,  andrea.addazi@infn.lngs.it} \\
{\it \it Dipartimento di Fisica,
 Universit\`a di L'Aquila, 67010 Coppito, AQ \\
LNGS, Laboratori Nazionali del Gran Sasso, 67010 Assergi AQ, Italy}
\end{center}

\vspace{1cm}
\begin{abstract}
\large

We discuss general aspects of non-relativistic quantum chaos theory of scattering 
of a quantum particle on a system of a large number of naked singularities. 
We define such a system {\it space-temporal Sinai billiard}
We discuss the problem in semiclassical approach. 
We show that in semiclassical regime the formation of  trapped periodic semiclassical orbits 
inside the system is unavoidable. 
This leads to general expression of survival probabilities 
and scattering time delays, 
expanded to the chaotic Pollicott-Ruelle resonances. 
Finally, we comment on possible generalizations
of these aspects 
to relativistic quantum field theory.

\end{abstract}

\baselineskip = 20pt

\section{Introduction}


The Quantum Chaos theory studies the systems 
in classical chaos and quantum mechanics regimes.
 Quantum Sinai Billiard is a well known example 
 of quantum chaotic system \cite{QuantumChaos1, QuantumChaos2}. 
 However, quantum chaotic scatterings  in 
 contest of general relativity were not studied as well
 in literature. 

In this paper, we will discuss general aspects of 
quantum scatterings of wave functions 
on a complicated space-time topology
composed of a large number of horizonless 
singularities, randomly oriented. 
We dub such a system {\it space-temporal Sinai billiard}
 What we will expect is that the initial probability will be fractioned into
two contributions. In fact, a part of the initial probability density will "escape" 
by the system while a part will remain "trapped" forever in the system 
because of back and fourth scatterings among the singular geometries. 
This can be easily understood by a classical chaotic mechanics point of view.
In fact, the definition of a classical chaotic scatterings  of a particle is the following:
a classical mechanics scattering problem in which 
the incident particle can be trapped ideally forever 
in a class of classical orbits; but the periodic orbits are
unstable saddle solutions and their number grows exponentially 
with time. 
Chaotic scatterings have a high sensitivity to the initial conditions
manifesting itself in a fractal chaotic invariant set, which is also 
called {\it chaotic saddle} \cite{QCS1,QCS2}.
Energy shells closed to the chaotic saddle energy shell
will continue to be chaotic. 
In simpler chaotic systems, examples are Kolmogorov-Arnold-Moser (KAM) elliptic islands, 
that contain stable periodic orbits.
KAM stable periodic orbits undergo to chaotic bifurcations, 
rupturing the smoothed topology of the invariant 
set \cite{QCS3,QCS4}. As usually happening for chaotic saddles, 
KAM islands are surrounded by a layer of chaotic trajectories. 
Another typical example is the hyperbolic set of hyperbolic unstable 
trajectories: solutions are exponentially growing or decreasing but the
number of directions are constants of motion. 
In our case,  periodic orbits will be 
forever trapped in back and forth scatterings
among the the space-temporal Sinai biliard. As generically happening in 
classical chaotic scatterings problems,
these trajectories will necessary exist in the phase space of the system
Our problem is nothing but a complication with respect to
a simpler and well known example of classical chaotic scattering 
problem: a 2d classical elastic scattering of
a particle on a system of N fixed disks of radius $a$
\cite{QCS5,QCS6}.
In this simple problem, kinetic energy is assumed to be conserved, 
{\it i.e} no any dissipations are considered. 
For one disk the problem is trivially un-chaotic:
the differential cross section is just $\frac{d\sigma}{d\theta}=\frac{a}{2}|\sin \frac{\theta}{2}|$
for $\theta$ in the range $[-\pi,\pi]$;
and no trapped periodic orbit are possible. 
However, with two disks, an unstable periodic orbit is the one 
bouncing back and forth forever among the two disks.
With the increasing of the number of disks one can easily 
get that the number of trapped periodic orbit will exponentially 
increase. 
For example, as shown in \cite{QCS7,QCS8}, 
in a three disks' system, the 
number of unstable periodic orbits proliferate as $2^{n}$
where $n$ is the number of bounces in unit of the period.
if the radius is the distance among the 
next neighboring disk is $R>2.04822142\, a$. 
From Classical chaotic scatterings
we can get the main feature of the quantum semiclassical 
chaotic problem associated and about semiclassical 
periodic orbits. 
So, 
because of multiple diffractions and back and fourth scatterings, 
one will also expect that the resultant wave function is "chaotized" by the system:
the total wave function is a superposition of the initial one 
plus all the spherical ones coming from each "scatterators".  
A part of the initial infalling information will be trapped "forever" 
in the system, {\it i.e} for all the system life-time.
In order to describe the evolution of the infalling informations,
a quantum mechanical approach 
based on wave functions  is not useful, in this system.
A wave functions approach can be substituted by a quantum statistical mechanics
approach in terms of density matrices. 
From the point of view of a Quantum field theory, a S-matrix approach is not useful in this case,
even if "fundamentally true":
in order to calculate $\langle in|S| out \rangle$ ($in$ is the in-going plane wave,
where $out$ is the out-going result), we have to 
get unknown informations on the precise geometric configuration 
inside the system and about the trapped information state inside it. 
Such a system can emit a quasi thermalized mixed information state
without losing any informations at fundamental level. 
In other words, we 
 suggest that the space-time non-trivial topology prepares an entangled 
state as well as an experimental apparatus can prepare 
an entangled state by an initial pure state. 
Then considering also possible interaction terms among quantum particles,
the chaotization effect will also be more dramatically efficient. 
In particular, 
in quantum field theory, interactions in the lagrangian density functional induce n-wave mixings 
inside the space-temporal Sinai billiard. Thinking about the ingoing state as a collection 
of coherent quantum fields, these will be scattered into 
the system and, they will meet each others inside "the trap", they will scatter each others,
coupled by lagrangian interactions.
A complicated cascade of hadronic and electromagnetic processes
is expected. For example, these will produce a large amount of neutral pions, 
that will electromagnetically decay into two entangled photons 
$\pi^{0}\rightarrow \gamma \gamma$ ($\tau\simeq 10^{-16}\, \rm s$ in the rest frame). 
However, also from only one plane wave infalling in the system, 
the final state emitted by the system will be a mixed state:
this is just an effect of the information losing inside the system
because of trapped chaotic zones inside.
This phenomena is a new form of quantum decoherence  
induced by the space-time topology. Usually, quantum decoherence is
the effective losing of infalling informations in a complex system,
like coherent light pumped in a non-linear crystal. 
In this case,  the complex topology of space-time  
catalyzes the effective losing of information. 

A possible applications of our result 
is in contest of theoretical cosmology. 
In particular, it was suggested the presence of topologically defects, as a net of cosmic strings, 
can affect the gaussianity of the CMB spectrum \cite{CS1,CS2,CS3,CS4,CS5}. 
A critical cosmic string sources 
a conic naked singular geometry \cite{CS1,CS2,CS3,CS4,CS5}. 
So that, a net of cosmic strings is 
thought as a complicate superposition of 
conic geometries. 

This paper is organized as follows: 
in section 2 we will discuss the chaotic scattering problem 
on the space-temporal Sinai Billiard in classical (subsection 2.1) and
semiclassical approaches (subsection 2.2), then we will discuss the problem
with non-relativistic quantum scattering methods (subsection 2.3). 
At the end of  section 2,  we will comment on possible
extension to the chaotic quantum field theory effects (subsection 2.4). 
Finally we show our conclusions and outlooks in section 3.

\section{Chaotic space-temporal Sinai Billiard}
In this section, 
we will discuss the non-relativistic quantum scattering problem 
of a wave function on a system of naked singular geometries, 
which we dub space-temporal Sinai Billiard. In particular, we are interested to 
conic singular geometries disposed in a idealized 3D box. 
We will discuss the general aspects of the problem in classical and semiclassical approach.
Then, we will discuss what happen in a 3D box of N conic singularities,
in the non-relativistic quantum scattering approach. 
Finally, we will comment on a generalization to quantum field theory
and quantum particle interactions. 

\subsection{Classical chaotic scattering on a Space-temporal Sinai Billiard}

The classical chaotic scattering of a particle on a Space-temporal Sinai Billiard 
is characterized by a classical Hamiltonian system 
$\dot{\vect{r}}=\partial H/\partial \vect{p}$ 
and $\dot{\vect{p}}=-\partial H/\partial \vect{r}$
with an initial condition $x_{0}=(\vect{r}_{0},\vect{p}_{0})$
in the space of phase. 
In particular, 
Let us consider the case of a conic singularity:
supposed disposed along the z-axis, the conic metric is
\be{metriccone}
ds^{2}=-dt^{2}+dr^{2}+\left(1-\frac{\Psi}{2\pi}\right)^{2}r^{2}d\psi^{2}+dz^{2}
\ee 
where $\Psi$ is the deficit angle, related to the opening angle 
as $\Theta=2\pi-\Psi$. 
To consider a generic field function  on a conic surface 
 is equivalent to consider these on a 
 Euclidean plane with an extra periodicity condition
\be{convenient1}\phi(\theta)=\phi(\theta+\Theta)\ee
where $\Theta$ is the open angle of the cone
 and $\theta$ is a new angle variable defined so that 
 \be{convenient2}
 ds^{2}_{E}=d\tau^{2}+dr^{2}+r^{2}d\theta^{2}+dz^{2}
\ee
 Such an Euclidean background has 
 a topology $S^{1}\times S^{1}\times R^{2}$ or $T^{2}\times R^{2}$,
 {\it i.e} is a cilinder with a torus as its base.

In a box of cones, we consider N conic singularities with random orientation of their axis. 
In particular, we can define N Hamiltonian 
systems for each cones, describing the
motion of the particle on each of N cones.
Clearly, one can obtain similar systems 
by the geodesic equations 
$\ddot{x}^{\mu}+\Gamma_{\alpha\beta}^{\mu} \dot{x}^{\alpha}\dot{x}^{\beta}=0$
of the particles in each conic metrics,
{\it i.e} the propagation of the particle 
on the conic hypersurfaces. 
We can easily show that the effective 
Hamiltonian obtained for the propagation 
on one cone 
is
\be{effective11}
H_{I}=\frac{1}{2m}p_{i}g^{ij}p_{j}=\frac{p^{2}}{2m},\,\,\,\,\,\,x_{t}(\theta+\Theta)=x_{t}(\theta)
\ee
where we have used 
(\ref{convenient1},\ref{convenient2}) (in Minkowskian form). 
This Hamiltonian is written choosing the reference frame as 
the z-axis oriented along the cone's axis. 
As a consequence, also with one cone, 
we have a class of trapped trajectories
infinitely going around the $\theta$-direction. 
Clearly, for a system of N cones randomly oriented,
extra angles parameterizing their axis
directions with respect to the chosen z-axis will
enter in the definition (\ref{effective11}).
The solution of such a system will be determined by 
a trajectory 
$x_{t}=\phi^{t}(x_{0})$ solving the Chauchy problem of classical mechanics.
In this case, we will expect a 
proliferation of trapped periodic unstable trajectories,
as anticipated in the introduction, 
because of an infinite back and forth 
scattering among the N cones. 

Let us define the action of the classical problem:
\be{action}
S(E)=-\int_{\Sigma} \vect{r}\cdot \vect{p}
\ee
where $\Sigma$ is the energy shell $H=E$
where a scattering orbit is sited. 
The time delay is defined 
as 
\be{time}
\mathcal{T}(E)=\frac{\partial S}{\partial E}
\ee
If the impact parameters of the initial orbits $\rho$ 
has a probability density $w(\rho)$, 
the probability density conditioned by energy $E$ of the corresponding 
time delays is 
\be{Prob}
P(\tau|E)=\int d\rho w(\rho)\delta(\tau+\mathcal{T}(\rho|E))
\ee
where the condition "corresponding time delays" is encoded
in the integral though the Dirac's delta. 
(\ref{Prob}) is useful to describe the escape of the particle from the 
trapped orbits' zone. 
Inspired by $N$ disks problems studied in literature \cite{QCS7,QCS8}, 
an hyperbolic invariant set is expected to occur.
In this case the decays' distribution rate is 
expected to exponentially decrease, {\it i.e}
\be{expdcr}
lim_{t\rightarrow \infty}\frac{P(\tau|E)}{t}=-\gamma(E)
\ee
On the other hand, for non-hyperbolic sets, 
like KAM elliptic islands, 
power low decays
are generically expected 
$P(t|E)\sim 1/t^{\alpha}$,
where $\alpha$ depends by 
the articular density of trapped orbits.

Now, let us discuss the time delays in our system of cones. 
If unstable periodic orbit exists in our scattering problem,
eq.(\ref{time}) will have $\rho$-poles,
{\it i.e} it becomes infinite for  
precise initial impact parameters $\rho$. 
Let starts with the simplest case
of a scattering on a simple cone.
In this case, 
the 
integral (\ref{time}) 
has only a couple of asymptotic divergent 
direction along the path 
$x_{t}^{\theta}=(\theta,p_{\theta})$.
The particle will be infinitely trapped in this 
path if and only if its 
initial incident direction is 
parallel to open angle $\theta$ of the cone.
This condition correspond to all trajectories
with a $z$ value in the range of the cone height. 

Now let us complicate the problem considering two cones.
In these case the number of divergent asymptotes of $\mathcal{T}$
correspond to three couples: 
i) cycles around the first cone, ii) cycles around the second cone,
iii) trapped back and forth trajectories between the two cones. 

One can easily get that for a N number of conic singularities 
the number of the divergent asymptotes 
for the time-delay function
will proliferate. 
These divergent asymptotes are connected to the fractal character 
of the invariant set. 
A geometric way to see the problem is the following:
we can consider a $2\nu-2$ Poincar\'e surface with section in the Hamiltonian
flown on a fixed energy surface, where $\nu$ is the number of degree of freedom
of the system. In our case, we consider a 4d Poincar\'e surface. 
The time-delay of the orbit necessary to go-out 
from the cones at large enough distances 
is $\mathcal{T}_{\pm}(\rho|E)$, for every initial 
Chauchy condition in the Poincar\'e section. 
 $\mathcal{T}_{+}(\rho|E)\rightarrow \infty$ for
 stable surfaces of orbits trapped forever.
 On the other hand, $\mathcal{T}_{-}\rightarrow \infty$
 on unstable manifolds of orbits 

 In other words, $|\mathcal{T}_{-}(\rho|E)|+|\mathcal{T}_{+}(\rho|E)|$
 is a localizator functions for the fractal set trapped trajectories.
 
Let us remind the definition of sensitivity to initial conditions,
defined by the Lyapunov exponents 
\be{Lyapunov}
\lambda(x_{0}|\delta x_{0})=lim_{t\rightarrow \infty}\frac{1}{t}\frac{|\delta x_{t}|}{|\delta x_{0}|}
\ee
where $\delta x_{0,t}$ are infinitesimal perturbation of the 
initial condition $x_{0}$ and the resultant 
orbit $x_{t}$. 
In general, the Lyapunov exponents depend on the initial perturbation 
and on the orbit perturbation.
However, $\lambda$ becomes un-sensible by the orbit
in ergodic invariant sets. 
These sets are characterized by 
the following hierarchy of Lyapunov exponents
in a system with $\nu$-degrees of freedom:
\be{lambdaf0}
0=\lambda_{\nu}\leq \lambda_{\nu-1} \leq .... \leq \lambda_{2} \leq \lambda_{1}
\ee
while
\be{lambdaf1}
0=\lambda_{\nu+1}\ge ... \ge \lambda_{2\nu}
\ee
In a Hamiltonian system, 
the symplectic flows of the Hamiltonian 
operator implies 
that 
$$\sum_{k=0}^{2\nu}\lambda_{k}=0$$
and 
$$\lambda_{2\nu-k+1}=-\lambda_{k}$$
where $k=1,2,...,2\nu$. 
In our case, the number of degree of freedom 
is $\nu=3$, so that the number of
independent Lyapunov's exponents characterizing 
the chaotic scattering is three. 

The exponentially growing number of unstable
periodic trajectories inside the invariant set 
is characterized by a topological number 
\be{topn}
h=lim_{t\rightarrow \infty}\frac{1}{t}ln(\mathcal{N}\{\tau_{o}\geq t\})
\ee
where $\mathcal{N}$ is the number of 
periodic orbits of period minor than $t$, $\tau_{o}$ is the periodic orbit time. 
Such a number is the so called topological entropy
$h>0$ if the system is chaotic
while $h=0$ if non-chaotic. 
For a system like 
a large box of cones, 
this number will be infinite. 
Such a number will diverge just with only three cones
as happen just in a system of three 2d-disks. 

In our system, as for disks, a  
 hyperbolic invariant set or something of similar is expected.  
For this set $\delta V$ small volumes
are exponentially 
stretched 
by  
\be{factorgrowing}
g_{\omega}=exp\{\sum_{\lambda_{k}>0}\lambda_{k}t_{\omega}\}>1
\ee
because of its unstable orbits; 
where $t_{\omega}$ is the time interval associated
to the periodic orbit of period $n$,
{\it i.e} to the symbolic 
dynamic $\omega=\omega_{1}....\omega_{n}$
corresponding to 
all the nonperiodic 
and periodic orbits 
remaining closed 
in a $\delta V$ for a time
$t_{\omega}$.
Using (\ref{factorgrowing}),
we can 
weight the probabilities for
trapped orbits as 
\be{probweight}
\mu_{\alpha}(\omega)=\frac{|g_{\omega}|^{-\alpha}}{\sum_{\omega}|g_{\omega}|^{-\alpha}}
\ee
This definition is intuitively understood:
a highly unstable
trajectory with $g_{\omega}>>1$ is 
weighted as $\mu_{\alpha}\simeq 0$.
The definition (\ref{probweight}) is normalized 
$\sum_{\omega}\mu_{\alpha}(\omega)=1$.
With $\alpha=1$ we recover the ergodic definition
for the Hamiltonian 
system

An intriguing question will be if we can determine the 
Hausdorff dimension of the fractal sets
for our box of cones. 
In principle, the answer is yes, but in practice 
the problem seems really hard to solve.
In order to get the problem let us define 
the Ruelle topological pressure 
 \be{Ruelle}
 P(\alpha)=lim_{t\rightarrow \infty}\frac{1}{t}ln \sum_{\omega, t<t_{\omega}<t+\Delta t}|g_{\omega}|^{-\alpha}
 \ee
Ruelle topological pressure is practically independent by $t,\Delta t$ for 
a large 
$\Delta t$. 
The Ruelle topological pressure has a series of useful relations:

1) $P(\alpha_{1}+\alpha_{2})\leq P(\alpha_{1})+P(\alpha_{2})$

2) $P(0)=h$, {\it i.e} for $\alpha=0$ the Ruelle topological pressure
is just equal to the topological entropy.

3) $P(1)=-\gamma$, {\it i.e} for $\alpha=1$ the Ruelle topological 
pressure is just equal to the escape rate.

4) The Ruelle topological pressure is connected to Lyapunov's exponents as
$$\frac{dP}{d\beta}(1)=-lim_{t\rightarrow \infty}\sum_{\omega,t<t_{\omega}<t+\Delta t}
\mu_{1}(\omega)ln|g_{\omega}|=-\sum_{\lambda_{k}>0}\lambda_{k}$$

The last relation is the one 
connecting the Ruelle topological pressure with the Hausdorff
dimension $d_{H}$: 
5) $P(d_{H})=0$.
The Hausdorff dimension of a system with $\nu$ d.o.f is 
bounded as $0\leq d_{H} \leq \nu-1$
for the subspace of unstable directions, 
while a corresponding set of stable directions 
has exactly the same dimension of the previous one. 
Let us note that for a system with $\nu=1$
the Hausdorff dimension will collapse
to $d_{H}=0$, {\it i.e} no chaotic dynamics. 
In our case, $0\leq d_{H} \leq 2$
and in principle it can be founded 
as a root of the Ruelle topological pressure. 

\subsection{Semiclassical chaotic scattering on a Space-time Sinai Biliard}

In Semiclassical approach, 
the main aspects of fully classical limit 
are not jeopardized by quantization: trapped periodic orbits, invariant sets
and so on. 
In semiclassical approach we can 
generalize the classical notion of time delay 
for a semiclassical quantum system. 

Let us remind, just to fix our conventions,  that $\psi_{t}(\vect{r})$ 
is obtained by an initial $\psi_{0}(\vect{r}_{0})$ 
by the unitary evolution 
\be{psitpsi}
\psi_{t}(\vect{r})=\int d\vect{r}_{0}K(\vect{r},\vect{r}_{0},t)\psi_{0}(\vect{r}_{0})
\ee
where $K$ is the propagator, represented as a non-relativistic Feynman path integral 
as 
\be{Krrz}
K(\vect{r},\vect{r}_{0},t)=\int \mathcal{D}\vect{r} e^{\frac{i}{\hbar}S}
\ee
where
$$I=\int_{0}^{t} dt L(\vect{r},\dot{\vect{r}})$$
I the action and $L$ the lagrangian of the particle. 
The semiclassical limit is obtained in the limit 
$$I=\int_{0}^{t}[\vect{p}\cdot d\vect{r}-Hd\tau]>>\hbar$$
so that the leading contribution to the path integral 
is just given by classical orbits. 
The corresponding WKB propagator
has a form 
\be{WKB}
K_{WKB}(\vect{r},\vect{r}_{0},t)\simeq \sum_{n}\mathcal{A}_{n}(\vect{r},\vect{r}_{0},t)e^{\frac{i}{\hbar}I_{n}}
\ee
where we sum on all over the classical orbits of the system.
The amplitudes $\mathcal{A}_{n}$
are 
\be{amplitudesss}
\mathcal{A}_{n}(\vect{r},\vect{r}_{0},t)=\frac{1}{(2\pi i \hbar)^{\nu/2}}\sqrt{|det[\partial_{\vect{r}_{0}}\partial_{\vect{r}_{0}}I_{n}[\vect{r},\vect{r}_{0},t]]|}e^{-\frac{i\pi h_{n}}{2}}
\ee
($h_{n}$ counts the number of conjugate points along the n-th orbit). 

The probability amplitude is related to Lyapunov exponents 
as 
\be{unstableAA}
|\mathcal{A}_{n}|\sim exp\left(-\frac{1}{2}\sum_{\lambda_{k}>0}\lambda_{k}t \right)
\ee
along unstable orbits.
On the other hand, 
\be{stableSSS}
|\mathcal{A}_{n}|\sim |t|^{-\nu/2}
\ee
along stable orbits

The level density of bounded quantum states 
is related to the trace of the propagator.
In semiclassical limit, 
the trace over the propagator
is peaked on 
around the periodic orbits 
and stationary saddles. 
This allows to quantize \'a l\'a Bohr-Sommerfeld
semiclassical unstable periodic orbits,
that are densely sited in the invariant set.
The 
semiclassical quantum time delay is 
\be{timed}
\mathcal{T}=\int \frac{d\Gamma_{ph}}{(2\pi\hbar)^{\nu-1}}[\delta(E-H_{0}+V)-\delta(E-H_{0})]+O(\hbar^{2-\nu})+2\sum_{p}\sum_{p} \tau_{a=1}^{\infty}\tau_{p}\frac{\cos \left(a\frac{S_{p}}{\hbar}-\frac{\pi a}{2}{\bf m}_{p}\right)}{\sqrt{|det(\mathcal{M}_{p}^{a})|}}+O(\hbar)
\ee
where $d\Gamma_{ph}=d\vect{p}d\vect{r}$.
The sum is on all over the periodic orbits, where primary periodic orbits are labelled as $p$ and the
number of their repetitions are labelled as $a$);
$S_{p}(E)=\int \vect{p}\cdot d\vect{r}$, $\tau_{p}=\int_{E}S_{p}(E)$,
${\bf m}_{p}$ is the Maslov index, 
and $\mathcal{M}$ is the $(2\nu-2)\times (2\nu-2)$  Poincar\'e map matrix defined in the 
neighborhood of the a-orbit.

Now, let us consider a simplified problem with only $\nu=2$
d.o.f, in order to more easily get analytical important proprieties 
of semiclassical chaotic scatterings and their features.
Let us consider a generic projection of our box of cones 
to a 2d plane. Now, we study the dynamics
in this plane, ignoring the existence of a third dimension.
However, we can be so general in our consideration 
to be practically valid for every chosen projection!
Clearly, we remark that we know well how this problem 
can be only a different simplified problem with respect 
the 3d one. 
In this case, the matrix $\mathcal{M}$ has two eigenvalues:
$\{g_{p},g_{p}^{-1} \}$, where $g_{p}$ is the classical factor
$|g_{p}|=exp(\lambda_{p}\tau_{p})$.
The complicate equation (\ref{timed})
for the time delay is just reduced 
to 
\be{TEred}
\mathcal{T}(E)=\mathcal{T}_{0}(E)-2\hbar Im\frac{d\,ln\,Z(E)}{dE}+O(\hbar)
\ee
where $\mathcal{T}_{0}(E)$ is the analytical part of the time-delay function
given by the first integral in (\ref{timed}),
while $Z(E)$ is the Zeta function 
\be{Zfu}
Z(E)=\prod_{p} \prod_{a=0}^{\infty}\left(1-e^{ia\phi_{p}}\frac{1}{g_{p}^{a}\sqrt{|g_{p}|}}\right)
\ee
where 
$$\phi_{p}=\frac{1}{\hbar}S_{p}-\frac{\pi}{2}{\bf m}_{p}$$
From (\ref{TEred}) and (\ref{Zfu}) one could get,
as an application of the Mittag-Leffler theorem, 
that 
the pole of the resolvent operators 
exactly corresponds to the zeros of the 
Zeta function. In complex energies' plane, 
the contribution of periodic orbits to 
the trace of the resolvent operator 
is related to the Z function by the simple relation
\be{relasimple}
tr \frac{1}{z-H}|_{p}= \frac{d}{dz}lnZ(z)=\frac{1}{i\hbar}\sum_{p}\sum_{a}\tau_{a}e^{ia\phi_{p}}\frac{1}{|g_{p}|^{a/2}}
\ee
(we omit extra higher inverse powers of $|g_{p}|$).
But the poles of the resolvent operator and the zeros of the Zeta function 
are nothing but scattering resonances:
$$Z(E_{a}=\mathcal{E}_{a}-i\Gamma_{a}/2)=0$$
Let us comment that if the invariant set contains
a single orbit, 
resonances $E_{a}$
satisfy the Bohr-Sommerfeld quantization condition
$$S_{p}(\mathcal{E}_{a})=2\pi \hbar \left( a+\frac{1}{4}{\bf m}_{p}\right) +O(\hbar^{2})$$
while widths satisfy
$$\Gamma_{a}=\frac{\hbar}{\tau_{p}}\,ln\,|g_{p}(\mathcal{E}_{r})|+O(\hbar)$$
This last relation is intuitively understood: 
for a large instability of the periodic orbit $g_{p}>>1$,
the resonances' lifetime $\tau_{a}=\hbar/\Gamma_{a}<<1$. 

Let us return on our general problem, from 2d to 3d. 
The resonances 
will not always dominate the time evolution
of a wavepacket.
In fact, in a system like our 
one, one could expect
so many resonances
that after the first decays 
the system will proceed
to an average distribution over these
resonances' peaks.  
Considering a wavepacket $\psi_{t}(\vect{r})$
over many resonances in a region $W$ in the $\nu$-dimensional space,
the quantum survival probability is 
\be{intovr}
P(t)=\int_{W}|\psi_{t}({\bf r})|^{2}d\vect{r}
\ee
that can be also rewritten in terms of the initial density 
operator $\rho_{0}=|\psi_{0}\rangle \langle \psi_{0}|$
as 
\be{intial}
P(t)=tr \mathcal{I}_{D}({\bf r})e^{-\frac{iHt}{\hbar}}\rho_{0}e^{+\frac{iHt}{\hbar}}
\ee
where $\mathcal{I}_{D}$ is a distribution equal to $1$ for ${\bf r}$
into $D$ while is zero out of the region $D$. 
As done for the time-delay, one can 
express the survival probability in a semiclassical form
\be{survsemicl}
P(t)\simeq \int \frac{d\Gamma_{ph}}{(2\pi \hbar)^{f}}\mathcal{I}_{D}e^{{\bf L}_{cl}t}\tilde{\rho}_{0}+O(\hbar^{-\nu+1})+\frac{1}{\pi \hbar}\int dE\sum_{p}\sum_{a}\frac{\cos\left(a\frac{S_{p}}{\hbar}-a\frac{\pi}{2}{\bf m}_{p} \right)}{\sqrt{|det({\bf m}_{p}^{a}-{\bf 1})|}}\int_{p}\mathcal{I}_{D}e^{{\bf L}_{cl}t}\tilde{\rho}_{0}dt +O(\hbar^{0})
\ee
 where ${\bf L}_{cl}$ is the classical Liouvillian operator,
 defined in terms of classical Poisson brackets as 
 ${\bf L}_{cl}=\{H_{cl},...\}_{Poisson}$; $\tilde{\rho}_{0}$
 is the Wigner transform 
 of the initial density state
 
 The Sturm-Liouville problem associated to ${\bf L}_{cl}$
 defines the Pollicott-Ruelle resonances 

 \be{LclSL}
 {\bf L}_{cl}\phi_{n}=\{H_{cl},\phi_{n} \}_{Poisson}=\lambda_{n}\phi_{n}
 \ee
The eigenstates $\phi_{n}$ are Gelfald-Schwartz distributions.
They are the ones with unstable manifolds in the invariant set. 
On the other hand, the adjoint problem 
 \be{LclSL}
 {\bf L}_{cl}^{\dagger}\tilde{\phi}_{n}=\tilde{\lambda}_{n}\tilde{\phi}_{n}
 \ee
has eigenstates associated to stable manifolds. 
The eigenvalues $\lambda_{n}$  
are in general complex.
They have a real part $Re(\lambda_{n})\leq 0$
because of they are associated to an ensamble bounded periodic orbits.
On the other hand $Im(\lambda_{n})$ describe 
the decays of the statistical ensambles.
As shown in 

one can expand 
the survival probability over the Pollicot-Ruelle resonances as 
\be{Pexp}
P(t)\simeq \int \sum_{n}\langle \mathcal{I}_{D}|\phi_{n}(E)\rangle \langle \tilde{\phi}_{n}(E)|e^{\lambda_{n}(E)t}| \phi_{n}(E)\rangle \langle \tilde{\phi}_{n}(E)|\tilde{\rho}_{0}\rangle 
\ee 
From this expansion, one can consider the 0-th leading order: 
it will be just proportional to an exponential $e^{\lambda_{0}(E)t}$.
The long-time decay of the system is
expected to be related to the classical escape rate $\gamma(E)$.
So that we conclude that the survival probability 
goes as $P(t)\sim e^{-\gamma(E)t}$, {\it i.e} $s_{0}=-\gamma(E)$. 

As a consequence, the cross sections from A to B $\sigma_{AB}=|S_{AB}|^{2}$
are dramatically controlled
by the Pollicott-Ruelle resonances. 
let us consider cross sections' autocorrelations
\be{autocc}
C_{E}(\bar{E})=\langle \sigma_{BA}(E-\frac{\bar{E}}{2}) \sigma_{AB}(E+\frac{\bar{E}}{2})\rangle-|\langle \sigma_{BA}(E)\rangle|^{2}
\ee 
with $E$ labelling the energy shell considered. 
Let us perform the Fourier transform 
\be{Ftcc}
\tilde{C}_{E}(t)=\int_{-\infty}^{+\infty}C_{E}(\bar{E})e^{-\frac{i}{\hbar}\bar{E}t}d\bar{E}
\ee
As done for the survival probability, we expand (\ref{Ftcc}) 
all over the Pollicott-Ruelle spectrum 
so that we obtain 
\be{CE}
\tilde{C}_{E}(t)\simeq \sum_{n} \tilde{C}_{n}exp(-Re\lambda_{n}(E)t)\cos Im \lambda_{n}(E)t
\ee
where $\tilde{C}_{n}$ are coefficients of this expansion. 
In particular the leading order of (\ref{CE}) is related to (\ref{Pexp})
for $Im \lambda_{0}=0$:
\be{related}
\tilde{C}_{E}(t)\simeq exp(-\gamma(E)t)
\ee
corresponding to the main Lorentzian peak 
\be{mainllrr}
C_{E}(\bar{E})\sim \frac{1}{\bar{E}^{2}+(\hbar \gamma(E))^{2}}
\ee
while (\ref{CE})
corresponds to 
a spectral correlation 
\be{summm}
C_{E}(\bar{E})\simeq \sum_{n}\left\{ \frac{C_{n}}{(\bar{E}-\hbar Im\lambda_{n})^{2}+(\hbar Re \lambda_{n})^{2}}+
 \frac{C_{n}}{(-\bar{E}-\hbar Im\lambda_{n})^{2}+(\hbar Re \lambda_{n})^{2}}\right\}
\ee

We conclude resuming that a semiclassical quantum chaotic scattering approach 
leads to following conclusions about the box of cones problem:
i) the existence of chaotic regions of trapped trajectories has to be a consequence 
of our scattering problem;
ii) the qualitative behavior of survival probability
and 
 correlation function is qualitatively understood as a decreasing
 function in time with an exponent determined by classical chaos scattering considerations.

\subsection{Non-Relativistic Quantum Scattering }

Let us consider the Schroedinger equation for a particle,
in a cone geometry \footnote{Perhaps this problem could be found in standard test of advanced quantum mechanics and  non-relativistic quantum scattering theory. I have not found any useful references about this particular problem 
of quantum scattering, so that I have just decided to repeat the exercise in all the details. 
 }.
\be{Hamiltonian}
i\frac{\partial}{\partial t}\psi(x)=-\frac{\Delta_{c}}{2m}+A\frac{\delta(r-\bar{r})}{r}
\ee
where $\Delta_{c}$ is the Laplacian in the conical geometry.
For simplicity, we have considered a cone with its axis coincident with the z-axis.
In fact, the radius of the cone boundary is $r=\bar{r}$, 
and it can be encoded in the equation as a 
$\delta$-potential, while $A$ is the dimensional "coupling" of the potential. 

As usually done for this type of problem, 
we can separate the variables
as
\be{phi}
\psi(t,x)\sim e^{-i\omega t}\phi_{n}(r)\left( \sin n\nu\theta, \cos n\nu\theta\right)^{T},\,\,\,\,\,n=0,1,2,...
\ee
and defining the adimensional parameter $a=2mA$
and
substituting (\ref{phi}) to (\ref{Hamiltonian}) we obtain 
\be{fnr}
\frac{d^{2}\phi_{n}(r)}{dr^{2}}+\frac{1}{r}\frac{d\phi_{n}(r)}{dr}+\left[k_{z}^{2}-\frac{n^{2}\nu^{2}}{r^{2}}-\frac{a}{r}\delta(r-\bar{r})\right]\phi_{n}(r)=0
\ee
We demand as contour conditions 
\be{fn}
\phi_{n}(a+o^{+})-\phi_{n}(a+o^{-})=0
\ee
so that we can map such a problem 
to another free-like equation
\be{free}
\frac{d^{2}\phi_{n}(r)}{dr^{2}}+\frac{1}{r}\frac{d\phi_{n}(r)}{dr}+\left(k_{z}^{2}-\frac{n^{2}\nu^{2}}{r^{2}}\right)f_{n}(r)=0
\ee
This equation can be also rewritten as 
\be{free}
\frac{d^{2}u_{n}(r)}{dr^{2}}+\left(k_{z}^{2}-\frac{n^{2}\nu^{2}}{r^{2}}\right)u_{n}(r)=0
\ee
where $u_{n}=r\phi_{n}$ and $k_{z}^{2}$.

The solution (regular) corresponding to the continuous part of the spectrum is 
\be{fnap}
\phi_{n}(r)=c_{n}^{0}J_{n \nu}(k_{z} r),\,\,\,\,\,\,r<\bar{r}
\ee
\be{fnam}
\phi_{n}(r)=c_{n}^{-}(k_{z})H_{n\nu}^{-}(k_{z} r)-c^{+}(k_{z})H_{n\nu}^{+}(k_{z} r),\,\,\,\,\,\, r>\bar{r}
\ee
These solutions are valid for all values of $a$ in the $\delta$-potential.
Our problem has two matching conditions 
\be{match1}
c^{0}_{n}(k_{z})J_{n\nu}(k_{z} \bar{r})=c^{-}_{n}(k_{z})H_{n\nu}^{-}(k_{z} \bar{r})-c^{+}_{n}(k_{z})H_{n\nu}^{+}(k_{z} \bar{r})
\ee
\be{match2}
c^{0}_{n}(k_{z})\left[ \frac{a}{k_{z} \bar{r}}J_{n\nu}(k_{z} \bar{r})+J'_{n\nu}(k_{z} \bar{r})\right]=c_{n}^{-}(k_{z})H'^{-}_{n\nu}(k_{z} \bar{r})-c_{n}^{+}(k_{z})H'^{+}_{n\nu}(k_{z} \bar{r})\ee
(prime is the differentiation with respect to the adimensional variable $k_{z} r$).

This problem can be viewed as a scattering one.
The corresponding solution for the S-matrix is 
\be{S}
S_{n}(k_{z})=\frac{a J_{n\nu}(k_{z} \bar{r})H_{n\nu}^{-}(k_{z} \bar{r})+2i/\pi}{a J_{n\nu}(k_{z} \bar{r})H_{n\nu}^{+}(k_{z}\bar{r})-2i/\pi}
\ee
related to $f_{n}$ as usual:
$$S_{n}=1+2ik_{z} f_{n}$$
so that 
$$|S_{n}|=1\rightarrow S_{n}=e^{2i\delta_{n}}$$
We also remind as $f_{n}$ is related to this phase $\delta_{n}$:
$$f_{n}=\frac{e^{2i\delta_{l}}-1}{2ik_{z}}=\frac{e^{i\delta_{n}}\sin \delta_{n}}{k_{z}}$$
Let us remind that, as usual, the asymptotic expansion of the radial part of the wave function 
can be written as the sum of the incident plane-wave on the conic geometry and 
the spherical one 
as 
$$\frac{1}{(2\pi)^{3/2}}\left[e^{ik_{z}z}+f(\theta,\phi)\frac{e^{ikr}}{r}\right]$$

Now, Let us consider a series of scatterings on a large number of N cones, 
disposed with a uniform random distribution of axis. 
Let us suppose a box of $n \times m \times p$ 
cones, $n$ in the x-axis, $m$ in y-axis, $p$ in z-axis
(not necessary disposed as a regular lattice).  
Let us call $\mathcal{N}_{1},\mathcal{N}_{2}$ the sides sited in the xy-planes, 
$\mathcal{M}_{1,2}$ in xz-planes, $\mathcal{P}_{1,2}$ in zy-planes, 
edges of the box of cones. 
Suppose an incident plane wave $\psi_{0}$
on the 2D surface $\mathcal{N}_{1}$,
with $n\times m$ cones:
$n\times m$ conic singularities will diffract the 
incident wave in $n\times m$-components.
We want to evaluate the S-matrix from the in-state 0 to the out-the box one. 
One will expect that a fraction of initial probability density
will escape from the box by the sides $\mathcal{N}_{1,2}\mathcal{M}_{1,2},\mathcal{P}_{1,2}$, 
 another fraction will be trapped "forever" (for a time-life equal to the one of the system) inside the box. 
As a consequence, we have to consider all possible diffraction stories/paths. 
We also have to consider more complicated diffraction paths: the initial wave
can scatter back and forth in the system before going-out.

We can consider the problem as a superposition of 
the initial wave function, assumed as a wave plane, and the diffracted wave functions for each 
conic singularities. In this system, we can label the position of all the conic singularities as 
$(i,j,k)$, where $i=1,..n$, $j=1,..,m$, $k=1,...,p$. 
The total wave function can be written as
\be{totalwave}
\phi_{0}+f(\vect{n}_{0},\vect{n}_{111})\frac{e^{ikr_{111}}}{r_{111}}+f(\vect{n}_{0},\vect{n}_{121})\frac{e^{ikr_{121}}}{r_{121}}+...+f(\vect{n}_{0},\vect{n}_{1N1})\frac{e^{ikr_{1N1}}}{r_{1N1}}
\ee
$$+f(\vect{n}_{111},\vect{n}_{121})\frac{e^{ikr_{121}}}{r_{121}}+...+f(\vect{n}_{111},\vect{n}_{1N1})\frac{e^{ikr_{11N}}}{r_{11N}}$$
$$+f(\vect{n}_{111},\vect{n}_{211})\frac{e^{ikr_{211}}}{r_{211}}+f(\vect{n}_{111},\vect{n}_{221})\frac{e^{ikr_{221}}}{r_{221}}+...+f(\vect{n}_{111},\vect{n}_{2M1})\frac{e^{ikr_{2M1}}}{r_{2M1}}+f(\vect{n}_{111},\vect{n}_{212})\frac{e^{ikr_{212}}}{r_{212}}$$
$$+...+f(\vect{n}_{111},\vect{n}_{21P})\frac{e^{ikr_{21P}}}{r_{21P}}+..+f(\vect{n}_{111},\vect{n}_{2MP})\frac{e^{ikr_{2MP}}}{r_{2MP}}+.....$$
where $\vect{n}_{0}$ is the wave versor of the incident plane wave, 
$\vect{n}_{ijk}$ are wave versors of the scattered waves from the conic singularities in positions $ijk$, 
$r_{ijk}$ are radii from positions $ijk$. 

Under this approximation, we can use the transition amplitudes 
of the one scattering problem considered in the previous section. 

The resultant wave function will be a superposition of an infinite series of waves. 
As a consequence, the total wave function will be highly chaotized by the superposition 
of all the scattered waves. 

An S-matrix for one possible diffraction path is
\be{Sone}
\langle in|S^{1th-short}|out\rangle=S_{0-111}S_{111-222}S_{222-333}...S_{(n-1)(m-1)(p-1)-(nmp)}
\ee
where $S_{111-222}$ represents the S-matrix for a process  from 
in-state (after a scattering on) $111$ and with an out-state (after a scattering on) $222$.
This formulation  can be consider if and only if the interdistances among singularities 
are much higher than the cones' sizes.

We can write a generic S-matrix for one diffraction path 
as 
\be{Sgeneric}
\langle in|S^{Kth}|out \rangle=S_{0-1jk}S_{ijk}S_{i'j'k'}.....S_{(i^{n-1}j^{m-1}k^{p-1})-(i^{n}j^{m}k^{p})}
\ee
 (\ref{Sgeneric})
with conditions 
\be{cond1}
i\leq i' \leq i+1
\ee
\be{ond2}
j\leq j' \leq j+1
\ee
\be{cond1}
k\leq k' \leq k+1
\ee
$$...$$
\be{cond1}
i^{n-1}\leq i^{n} \leq i^{n-1}+1
\ee
\be{cond2}
j^{m-1}\leq j^{m} \leq j^{m-1}+1
\ee
\be{cond1}
k^{p-1}\leq k^{p} \leq k^{p-1}+1
\ee
represent a class of paths 
similar to (\ref{Sone}). 

These class of paths  are "minimal" ones:
 there are not 
back-transitions. 
 "Minimal paths" are $n\times m \times p \times (n-1)$;
 while the number of non-minimal paths 
 will diverge. 

The total S-matrix
is the (infinite) sum on all diffraction paths
\be{TOT}
\langle in|S^{OUT}_{n}|out \rangle=\sum_{paths}\langle in |S^{K-th}_{n}|out \rangle
\ee

The S-matrix for one diffraction path cn be written as 
\be{explicitSpath}
\left(S^{Kth}\right)_{n}=\prod_{j=first}^{last}\frac{a_{j}J_{n\nu}(k_{j}\bar{r}_{j})H_{n\nu}^{-}(k_{j}\bar{r}_{j})+\frac{2i}{\pi}}{a_{j}J_{n\nu}(k_{j}\bar{r}_{j})H_{n\nu}^{+}(k_{j}\bar{r}_{j})-\frac{2i}{\pi}}
\ee
where the product is performed from the first scattering to the last one, 
and $a_{j},\bar{r}_{j},k_{j}$ depend by the particular j-th conic singularity
($k_{j}$ depends on the direction of the conic axis).

\subsection{Quantum field theories}

In this section we will formally discuss the problem 
of scattering from a QFT point of view.
 If one considers the path integral behavior in the UV energy regime,
 the fields' configurations start to "feel" the 
effect of the non-trivial topology 
and naked cones' singularities. 
Information is chaotically mixed in this limit. 
In fact fields start to be randomly diffused by 
presence of randomly oriented cones.
A part of the fields' energy density will be trapped in 
the irregularities. 

Suppose interdistances much higher than cones' dimensions.
This case is a simplified one with respect to the realistic problem. 
In this case, we can define a transition amplitude for each cone.
Let us suppose to be interested to calculate the transition amplitude for 
a field configuration $\phi_{0}$ to a field configuration $\phi_{N}$.
$\phi_{0}$ is the initial field configuration defined on a $t_{0}$, 
before entering in the system, while $\phi_{N}$ is a field configuration 
of a time $t_{N}$, corresponding to a an out-going state from the
system.
For simplicity, we can formalize the simplified problem as a 4D-box,
with $n \times m \times p$ 
conic singularities in 3D, $n$ in the x-axis, $m$ in y-axis, $p$ in z-axis
(not necessary disposed as a regular lattice).  
Let us call $\mathcal{N}_{1},\mathcal{N}_{2}$ the sides sited in the xy-planes, 
$\mathcal{M}_{1,2}$ in xz-planes, $\mathcal{P}_{1,2}$ in zy-planes, 
delimiting the 3D-space-box. 
Let us consider an incident field $\phi_{0}$
on the 2D plane $\mathcal{N}_{1}$,
with $n\times m$ conic singularities. 
Then the $n\times m$ conic singularities will scatter the 
incident field in $n\times m$-waves.
From each diffractions, the out-waves will scatter on a successive 
cones, penetrating in the box, 
or to the other nodes in the same plane $\mathcal{N}_{1}$,
and so on. Our problem is to evaluate the S-matrix from the in-state 0 to the out-the box state. 
One will expect that a fraction of initial probability density
will escape from the 3D box by the sides $\mathcal{N}_{1,2}\mathcal{M}_{1,2},\mathcal{P}_{1,2}$, 
 another fraction will be trapped "forever" (for a time-life equal to the one of the system) inside the box. 
As a consequence, one has to consider all possible diffraction stories or diffraction paths. 
Clearly, one has also to consider paths in which the initial wave
goes back and forth in the system before going-out.

One example of propagation Path $0-111-222-333-...-nmp-N$ 
\be{examplepath}
\langle\phi_{0},t_{0}|\phi_{111,in},t_{111,in}\rangle \langle\phi_{111,in},t_{111,in}|\phi_{111,out},t_{111,out}\rangle   
\langle\phi_{111,out},t_{111,out}|\phi_{222,in},t_{222,in}\rangle
\ee
$$\times \langle\phi_{222,in},t_{222,in}|\phi_{222,out},t_{222,out}\rangle...\langle\phi_{(n-1,m-1,p-1)},t_{(n-1),(m-1),(p-1)}|\phi_{nmp},t_{nmp}\rangle\langle\phi_{n,m,p},t_{n,m,p}|\phi_{N},t_{N}\rangle$$
where $|\phi_{ijk,in},t_{ijk,in}\rangle$ and $|\phi_{ijk,out},t_{ijk,out}\rangle$
are states before and after entering in the conic geometry $ijk$. 
In order to evaluate $\langle \phi_{0},t_{0}|\phi_{nmp},t_{nmp}\rangle$  
 one has to consider all the possible propagation paths 
from the initial position to the $nmp$-th conic singularity.
We define these amplitudes as
\be{MinkoP}
\langle \phi_{ijk},t_{ijk}|\phi_{i'j'k',in},t_{i'j'k',in} \rangle= \int_{\mathcal{M}_{0}}\mathcal{D}\phi e^{iI[\phi]}
\ee
while 
\be{MinkoP}
\langle \phi_{ijk,in},t_{ijk,in}|\phi_{ijk,out},t_{ijk,out} \rangle= \int_{\mathcal{M}_{ijk}}\mathcal{D}\phi e^{iI[\phi]}
\ee
where $\mathcal{M}_{0}$ is the Minkowski space-time, while $\mathcal{M}_{ijk}$ is the ijk-cone space-time. 
Again one can easily get that for a large system of naked conic singularities, 
it will exist a class of propagators' paths,
reaching the out state $|\phi_{N}, t_{N} \rangle$ 
only for a time $t_{N}\rightarrow \infty$.
A simple example can be the propagator paths
\be{propath}
|\langle \phi_{ijk},t_{ijk}|\phi_{i'j'k'},t_{i'j'k'} \rangle|^{2} |\langle \phi_{ijk},t^{(1)}_{ijk}|\phi_{i'j'k'},t_{i'j'k'}^{(1)} \rangle|^{2} ....|\langle \phi_{ijk},t^{(\infty)}_{ijk}|\phi_{i'j'k'},t_{i'j'k'}^{(\infty)} \rangle|^{2}
\ee
where $t_{ijk}^{\infty}>....>t_{ijk}^{(1)}>t_{ijk}$ and $t_{i'j'k'}^{\infty}>....>t_{i'j'k'}^{(1)}>t_{i'j'k'}$.
This amplitude is non-vanishing in such a system as an infinite sample of other ones. 
We can formally group these propagators in 
a $\langle BOX|BOX \rangle$ propagator, evaluating the probability 
that a field will remain in the box of cones after a time 
larger than the system life-time. 
On the other hand, let call $\langle BOX|OUT \rangle$
and $\langle OUT|OUT \rangle$
the other processes. 

Considering interactions, one will also use S-matrices. 
We can write a generic S-matrix for one diffraction path 
as \be{Sgeneric}
\langle in|S^{Kth}|out \rangle=S_{0-1jk}S_{ijk}S_{i'j'k'}.....S_{(i^{n-1}j^{m-1}k^{p-1})-(i^{n}j^{m}k^{p})}
\ee
A class of paths like the one in (\ref{Sone}), 
are like (\ref{Sgeneric})
with conditions 
\be{cond1}
i\leq i' \leq i+1
\ee
\be{cond2}
j\leq j' \leq j+1
\ee
\be{cond1}
k\leq k' \leq k+1
\ee
$$...$$
\be{cond1}
i^{n-1}\leq i^{n} \leq i^{n-1}+1
\ee
\be{cond2}
j^{m-1}\leq j^{m} \leq j^{m-1}+1
\ee
\be{cond1}
k^{p-1}\leq k^{p} \leq k^{p-1}+1
\ee
We call these class of paths "minimal paths". 
In fact, in these paths there are not 
back-transitions. 
The total number of "minimal paths" is 
is $n\times m \times p \times (n-1)$.
On the other hand, the number of paths 
with back and forth scatterings will diverge. 

As a consequence, the total S-matrix
is the sum over all possible infinite diffraction paths
\be{TOT}
\langle in|S^{OUT}_{n}|out \rangle=\sum_{paths}\langle in |S^{K-th}_{n}|out \rangle
\ee
accounting for all the paths leading from the in-state to 
the out-of-box state.

For a completeness of our discussion, 
let us reformulate the non-relativistic quantum problem 
in a non-relativistic path integral formulation.
We will use here the bracket-notation,
in which the propagator from $(x_{0},t_{0})$ to $(x_{1},t_{1})$ is
$$K(x_{0},t_{0};x,t_{1})=\langle x_{0},t_{0}| x_{1}, t_{1} \rangle$$
 
This will be equivalent to wave functions' formulation considered 
in section 3.2. 
In this case, a problem of $\langle OUT|OUT \rangle$ 
is reformulated not with propagators in the fields' space 
but in the same space-time points.
$\langle OUT|OUT \rangle$ will account for all possible 
paths leading to an in-coming state $|x_{0},t_{0}\rangle$
 to another state out of the box. 
 Again, such a problem is chaotized by the fact that 
 one has to consider the interference of all possible paths 
 passing for all possible conic geometries. 
 A simple example of a path inside the OUT-OUT class of paths 
 is $0-111-222-333-...-nmp-N$ 
\be{examplepath}
\langle x_{0},t_{0}|x_{111,in},t_{111,in}\rangle \langle x_{111,in},t_{111,in}|x_{111,out},t_{111,out}\rangle   
\langle x_{111,out},t_{111,out}|x_{222,in},t_{222,in}\rangle
\ee
$$\times \langle x_{222,in},t_{222,in}|x_{222,out},t_{222,out}\rangle...\langle x_{(n-1,m-1,p-1)},t_{(n-1),(m-1),(p-1)}|x_{nmp},t_{nmp}\rangle$$
where $|x_{ijk,in},t_{ijk,in}\rangle$ and $|x_{ijk,out},t_{ijk,out}\rangle$
are states before and after entering in the conic geometry $ijk$. 

One can find trapped propagators 
like 
\be{propath}
|\langle x_{ijk},t_{ijk}|x_{i'j'k'},t_{i'j'k'} \rangle|^{2} |\langle x_{ijk},t^{(1)}_{ijk}|x_{i'j'k'},t_{i'j'k'}^{(1)} \rangle|^{2} ....|\langle x_{ijk},t^{(\infty)}_{ijk}|x_{i'j'k'},t_{i'j'k'}^{(\infty)} \rangle|^{2}
\ee
where $t_{ijk}^{\infty}>....>t_{ijk}^{(1)}>t_{ijk}$ and $t_{i'j'k'}^{\infty}>....>t_{i'j'k'}^{(1)}>t_{i'j'k'}$.
A class of paths from OUT to BOX state will be attracted in these 
trapped paths. This is a reformulation of what we have concluded in Section 3.2. 


Now, let us return to QFT formulation. 
We are against a chaotic quantum field theory problem.
In a chaotic quantum field theory, 
there are not trapped trajectories in space-time 
but there trapped configurations in the 
infinite dimensional space of fields!
In analogy to semiclassical chaotic non-relativistic quantum mechanics, 
one can consider a semiclassical approximation in a regime in which 
the fields' action is much higher than $\hbar$:
$I>>\hbar$. 
In this approximation, we have a formal understanding of the 
chaotic quantum field theory problem. 
The corresponding WKB propagator
for a quantum field has a form 
\be{WKB}
\langle \phi_{0},t_{0}| \phi_{1},t_{1}\rangle\simeq \sum_{n}\mathcal{A}_{n}(\phi_{0},t_{0}|\phi_{1},t_{1})e^{\frac{i}{\hbar}I_{n}}
\ee
where we are summing on all over the classical orbits 
in the fields' configurations' space, 
while amplitudes $\mathcal{A}_{n}$
are 
\be{amplitudesss}
\mathcal{A}_{n}(\phi_{0},t_{0}|\phi_{1},t_{1})=\frac{1}{(2\pi i \hbar)^{\nu/2}}\sqrt{|det[\partial_{\phi_{0}}\partial_{\phi_{0}}I_{n}[\vect{r},\vect{r}_{0},t]]|}e^{-\frac{i\pi h_{n}}{2}}
\ee
where $h_{n}$ counts the number of conjugate points along the n-th orbits. 

In my knowledge, I will expect that all rigorous results obtained in literature 
of classical chaotic scatterings, about the existence of invariant set 
with their topological robust proprieties discussed in part above, 
are not rigorously extended for an infinite dimensional space of fields
and a complete theory regarding these aspects in QFT is not known to me.
Neverthless let us intuitively think something similar
 happen in space of fields, even if more complicated.
 The presence of chaotic zones of trapped periodic fields' configurations 
in a subregion of the configurations' space, corresponding 
to the one confined into our system, is expected for our problem.
Also for fields, chaotic unstable trajectories in the fields' space
are expected, as well as a large number of fields' resonances
in QFT S-matrices, generalizing Pollicot-Ruelle ones. 
The survival probability for a field are expected to 
exponentially decrease as in semiclassical quantum mechanical case. 

On the other hand, a general space of different fields,
the presence of interaction terms in the lagrangian leads
to tree-level transitions' processes
that has to be considered as leading orders in the
semiclassical saddle point perturbative expansion. 
As a consequence, chaotic fields' trapped trajectories
have to be thought as a multifields' ones. 
The result can be imagined as a 
chaotic cascade of processes among fields,
in which a part of different fields are trapped 
in the system continuosly interacting
and scatterings and decaying each others.
For example, let us imagine one pure electromagnetic wave
entering inside the box of cones. 
This starts to be diffracted into 
different direction,
so that initial coherent photons 
will start to re-meet each other in 
a different state. 
Of course, if their energy is enough, 
they can produce couples of $e^{+}e^{-}, q\bar{q}$ and so on. 
Then, these fields will interact each other through 
electromagnetic, strong and weak interactions.
The final system will be full of new fields, 
and it will have highly chaotic trapped zone.

\section{Conclusions and outlooks}

In this paper, we have discussed aspects of quantum chaos in a problem of scattering of 
quantum particles on a system of a large number of horizonless naked singularities, randomly oriented. 
We have assumed that the scattered particle has a too small mass to induce a relevant back-reaction 
to the space-time metric. 
The general results that we obtained can be resume as follows:

-In semiclassical regime, chaotic trapped zone of semiclassical periodic orbits will be inevitably formed 
inside the system. The initial information state transits to a highly chaotized final state.
The final state is fractioned in a forever trapped part inside the system -or at least trapped until the life-time of the 
space-time configuration. 

-The transition or survival probabilities calculated inside the system are dominated by the presence 
of a peculiar spectrum of resonances known in literature as Pollicott-Ruelle eigenvalues. 

We conclude that the problem considered can be only an example of a new interesting 
physics regime in which chaotic effects, quantum mechanics and general relativity 
cannot be neglected at the same time. 
In fact, the space-temporal Sinai billiard cannot be considered as a classical (gravitational) newtonian system
while it induces chaotic behaviors to quantum particles scattered on it. 
So that, the space-temporal Sinai billiard is an example of a quantum chaos problem in General Relativity. 
Finally, we mention that in recent papers, we suggested that this regime can lead to a reinterpretation of the quantum black hole 
nature and its information paradoxes. In particular, we suggested that black holes are a superposition 
of a large number of horizonless naked singularities \cite{Addazi:2015gna,Addazi:2015hpa,Addazi:2015cho}. 
In other words, the Penrose diagram of a black hole is an approximate superposition of a large number of Penrose diagrams of singular geometries. 
But we have to admit that our reinterpretation of black holes remains speculative and not enough quantitative. 
We hope that future progresses on quantum chaos in general relativity could be helpful
in order to understand black hole physics.

\vspace{1cm} 


{\large \bf Acknowledgments} 
\vspace{3mm}

I also would like to thank organizers of 14th Marcell Grossmann meeting in Rome and Karl Schwarzschild meeting 2015 in Frankfurt 
and LMU for hospitality during the preparation of this paper.  My work was supported in part by the MIUR research grant Theoretical Astroparticle Physics PRIN 2012CPPYP7 and by SdC Progetto speciale Multiasse La Societa` della Conoscenza in Abruzzo PO FSE Abruzzo 2007-2013.


\vspace{0.5cm}
\begin{thebibliography}{99}
  
  
  \bibitem{QuantumChaos1}
F. Haake,   {\it Quantum Signatures of Chaos} Edition: 2, Springer, 2001, ISBN 3-540-67723-2, ISBN 978-3-540-67723-9.

\bibitem{QuantumChaos2}
M.Berry, {\it Quantum Chaology}, pp 104-5 of {\it Quantum: a guide for the perplexed} by J-Al Khalili.


\bibitem{QCS1}
T. T\'el, M. Gruinz, {\it Chaotic Dynamics}, Cambridge University Press (2006), Cambridge UK.

\bibitem{QCS2}
J.M. Seoane and M. A. F. Sanju\'an 2013 Rep. Prog. Phys. {\bf 76} 016001.

\bibitem{QCS3}
V.I. Arnold, Russ.Math.Surv. 18:6:85-191. 
\bibitem{QCS4}
J.K.Moser, Mem.Am.Math.Soc. {\bf 81}: 1-60.

\bibitem{QCS5}
B.Eckhardt, J.Phys.A: Math. Gen. {\bf 20}: 5971-5979.
\bibitem{QCS6}
P.Gaspard and S.A. Rice, J.Chem.Phys. {\bf 90}:2225-2241. 
\bibitem{QCS7}
P.Gaspard, {\it Chaos, Scattering and Statistical Mechanics},
Cambridge University Press, Cambridge UK. 

\bibitem{QCS8}
K.T. Hansen, Nonlinearity {\bf 6}: 753-770.

\bibitem{QCS9}
M.M. Sano, J.Phys. A: Math. Gen. {\bf 27}: 4791-4803. 

\bibitem{CS1}
  H. B. Nielsen and P. Olesen, Nucl.Phys. {\bf B61}, 45 (1973). 
  
  \bibitem{CS2}
A. Vilenkin, Phys.Rev. D{\bf 23}, 852 (1981).
  
  \bibitem{CS3}
  P. Laguna and D. Garfinkle, Phys.Rev. D40, 1011 (1989). 
  
  \bibitem{CS4}
M. Christensen, A. L. Larsen, and Y. Verbin, Phys.Rev. {\bf D60}, 125012 (1999), arXiv:gr-qc/9904049 [gr-qc]. 
  
  \bibitem{CS5}
  M. E. Ortiz, Phys.Rev. {\bf D43}, 2521 (1991).
          
\bibitem{Addazi:2015gna} 
  A.~Addazi,
  arXiv:1508.04054 [gr-qc].

\bibitem{Addazi:2015hpa}
A. Addazi, arXiv:1510.05876 [gr-qc], to appear in KSM 2015,  C15-07-20.2.

\bibitem{Addazi:2015cho}
 A.~Addazi,
 Electron.\ J.\ Theor.\ Phys.\  {\bf 12}, no. 34, 0 (2015),  arXiv:1510.09128 [gr-qc].
  

\end{thebibliography}
\end{document}